\documentclass[reprint,showpacs,
 amsmath,amssymb,
 aps,pre
]{revtex4-1}

\usepackage{graphicx}
\usepackage{dcolumn}
\usepackage{bm}
\usepackage{color}

\begin{document}

\preprint{APS/123-QED}

\title{Cooperation dynamics in the networked geometric Brownian motion}
\author{Viktor Stojkoski$^{1}$}
\email{vstojkoski@manu.edu.mk}
\author{Zoran Utkovski$^{2,3}$}
\author{Lasko Basnarkov$^{1,4}$}
\author{Ljupco Kocarev$^{1,3}$}

\affiliation{
$^{1}$Academy of Sciences and Arts of the Republic of North Macedonia, P.O. Box 428, 1000 Skopje, North Macedonia}%
\affiliation{
$^{2}$Fraunhofer Heinrich Hertz Institute, Einsteinufer 37, 10587, Berlin, Germany}
\affiliation{
$^{43}$Faculty of Computer Science, University Goce Delcev Stip, P.O. Box 10-A, 2000 Shtip 2000, North Macedonia\\}%
\affiliation{
$^{4}$Faculty of Computer Science and Engineering, Ss. Cyril and Methodius University, P.O. Box 393, 1000 Skopje, North Macedonia\\}%

\date{\today}

\begin{abstract}
Recent works suggest that pooling and sharing may constitute a fundamental mechanism for the evolution of cooperation in well-mixed fluctuating environments. The rationale is that, by reducing the amplitude of fluctuations, pooling and sharing increases the steady-state growth rate at which individuals self-reproduce. However, in reality interactions are seldom realized in a well-mixed structure, and the underlying topology is in general described by a complex network. Motivated by this observation, we investigate the role of the network structure on the cooperative dynamics in fluctuating environments, by developing a model for \textit{networked} pooling and sharing of resources undergoing a geometric Brownian motion. The study reveals that, while in general cooperation increases the individual steady state growth rates (i.e. is evolutionary advantageous), the interplay with the network structure may yield large discrepancies in the observed individual resource endowments. We comment possible biological and social implications and discuss relations to econophysics.

\end{abstract}

\maketitle

\section{Introduction}

Cooperation has played a fundamental role in the evolution of systems consisting of individuals with different levels of complexity, ranging from simple cell to complex human behavior~\cite{Axelrod-2006}. However, natural selection imposes competition and thus the emergence of cooperation is predicated on the co-occurrence of a specific mechanism within the studied network of contacts~\cite{Nowak-2006five}.

A standard approach for examining the effect of different mechanisms on the cooperation dynamics in complex networks is through evolutionary graph theory~\cite{lieberman2005evolutionary}. Under this setting, the individuals interacting in a network are given a set of strategies which they can choose from, and a set of payoffs (changes in the individual resource endowment) that result from interactions with other individuals and their chosen strategies. In the simplest situation, each individual can either be a cooperator or a defector. A cooperator is someone who sacrifices its own resources in order to achieve a better collective performance, whereas defectors are individuals who exploit this cooperative behavior.

Since the pioneering works of Axelrod~\cite{Axelrod-2006}, and later Nowak et al.~\cite{Nowak-1993,Nowak-2005,Nowak-2006five,allen2017evolutionary}, on matrix games, i.e. pairwise interactions between individuals, a lot of effort has been put into uncovering the mechanisms required for cooperators to survive the invasion of defectors in networked societies. In particular, more general forms of interaction structures which capture group interactions have been discussed in~\cite{perc2017statistical,perc2013evolutionary,santos2008social}. In this context, it has been found that the introduction of spatial randomness represented by  heterogeneous resource endowments between individuals may unconditionally facilitate the evolution of cooperation~\cite{kun2013resource,mcavoy2015asymmetric}.

Despite the abundance of studies which capture such spatial stochasticity, a ubiquitous, yet largely unexplored scenario remains the one of cooperative interactions on complex networks in \textit{fluctuating environments} -- where the temporal evolution of resource endowments is strongly affected by their relative growth. In such situations, fluctuations have a net-negative effect on the time-averages, although having no effect on the ensemble (spatial) properties~\cite{peters2013ergodicity}. This observation, which is a result of the non-ergodicity of the fluctuation-generating process~\cite{peters2013ergodicity,peters2016evaluating}, yields evolutionary behavior which essentially differs from the one observed in standard models~\cite{radicchi2018uncertainty,stollmeier2018unfair}.

On this basis, it has been hypothesized that repeated pooling and sharing of resources which previously exhibit a fluctuating growth may constitute a fundamental mechanism for the evolution of cooperation in a well-mixed population. The rationale is that, by reducing the amplitude of fluctuations, pooling and sharing increases the steady state growth rate at which the individual cooperating entities self-reproduce~\cite{yaari2010cooperation,liebmann2017sharing,peters2015evolutionary}. A crucial real-life observation is, however, that interactions between individuals are  seldom  realized  in  a well-mixed structure, and they are instead driven by a complex network of contacts~\cite{allen2017evolutionary}.

Motivated by this observation, here we investigate the impact the \textit{complex network} topology on the cooperative dynamics in fluctuating environments, with networked individuals performing pooling and sharing of resources undergoing a geometric Brownian Motion (GBM). The noisy resource growth produced by GBM is a common model for  fluctuations~\cite{stollmeier2018unfair,zheng2018environmental,cvijovic2015fate}. The interactions are modeled by considering each individual to also be a pool through which its (direct) neighbors share resources. The model is evaluated analytically and numerically on four types of random graphs: random d-regular graph (RR)~\cite{bollobas2013modern}, Erdos-Renyi Poisson graph (ER)~\cite{erdos1960evolution}, Watts-Strogatz small-world network (WS)~\cite{watts1998collective} and Barabasi-Albert scale-free network (BA)~\cite{barabasi1999emergence}. Our findings suggest that, while there remains the general trend that cooperation increases the steady state growth rate of each individual (i.e. is evolutionary advantageous), the unique interplay between the non-ergodic fluctuation-generating process and the network topology may generate large discrepancies in the resource endowments. When present, this inequality has a negative effect on the growth rates of the individual entities, hampering their evolutionary performance. Parallels can be made to current societal discussions on wealth inequality~\cite{sciam2018}. 

The remaining of the paper is structured as follows. In Section~\ref{sec:model}, we describe the system model by providing details about the pooling and sharing mechanism, the networked interactions, and the properties of GBM. In Section~\ref{sec:analytical_results} we provide analytical results for the growth rate and the steady state behavior of the individual resource endowments. In Section~\ref{sec:numerical_results} we perform numerical experiments and comparison with the analytical results derived in the previous section. Finally, in Section~\ref{sec:discussion} we discuss our findings and give directions for future work. Some additional technical details are provided in the Appendix.

\section{Model}
\label{sec:model}

\subsection{Preliminaries}

Formally, we assume that there is a population of non-cooperative individuals, where the dynamics of resources $y_i(t)$ of each individual $i$ at time $t$ follow a geometric Brownian motion (GBM),
\begin{align}
\mathrm{d} y_i &= y_i \left( \mu \mathrm{d}t + \sigma \mathrm{d}W_i \right),
\label{eq:gbm}
\end{align}
with $\mu$ being the drift term, $\sigma$ the noise amplitude, and $\mathrm{d}W_i$ is an independent Wiener increment, $W_i(t) =\int_0^t \mathrm{d}W_i$. Without noise ($\sigma = 0$), the model is simply exponential growth at rate $\mu$. With $\sigma \neq 0$ it can be interpreted as exponential growth with a fluctuating growth rate. 

The advantage of modelling through GBM lies in its universality, as it represents an attractor of more complex models that exhibit multiplicative growth~\cite{aitchison1957lognormal,redner1990random}. 
Its non-ergodicity manifests as the difference between the growth rate observed in an individual trajectory and the ensemble average growth~\cite{peters2013ergodicity, peters2016evaluating}. 
In particular, the estimator for the growth rate, $g_i(y_i(t), t)$, of a single GBM trajectory is defined as
\begin{align}
  g_i(y_i(t), t) &= \frac{1}{t}\log\left(\frac{y_i(t)}{y_i(0)}\right),
\end{align}
where $y_i(0)$ is the initial condition. For simplicity, we assume $y_i(0)=1$. 

The time-averaged growth rate is found by letting time remove the stochasticity in the process, i.e., taking the limit as $t \to \infty$, which results in
\begin{align}
\lim_{t \to \infty} g_i(y_i(t), t) &= \mu - \frac{\sigma^2}{2}.
\label{eq:time-average-growth}
\end{align}
The ensemble growth rate, on the other hand, is found by substituting $y_i(t)$ with the average $\langle y \rangle$ of an infinite ensemble, where $\langle \cdot \rangle$ is the averaging operation. In other words, one lets the spatial dimension remove the stochasticity by averaging across all possible realizations. Mathematically, the solution is
\begin{align}
\lim_{N \to \infty} g_i(\langle y \rangle, t) &= \mu,
\label{eq:ensemble-average-growth}
\end{align}
where $N$ is the ensemble size.

If only a single system is to be modeled, in steady state only the time-averaged growth rate, Eq.~(\ref{eq:time-average-growth}), is observed. As discussed in~\cite{peters2013ergodicity}, the ensemble average growth rate~(\ref{eq:ensemble-average-growth}) is fictive, as it assumes averaging over ``imagined parallel universes''. Hence, in reality, it is the time-averaged growth rate that determines the evolutionary performance of an individual GBM trajectory. Simultaneously, it provides parallels to real-life phenomena. For instance, in evolutionary games the time-averaged growth rate is the geometric mean fitness for the accumulated payoff (resources) of a particular phenotype~\cite{saether2015concept}. In economic decision theory, where wealth (resources) dynamics follows a multiplicative process, the same growth observable arises naturally as the unique utility measure~\cite{peters2016evaluating}.

\subsection{Pooling and sharing of resources}
\textcolor{blue}{From an evolutionary perspective, individuals with lower noise amplitude should exhibit higher steady state growth rates and should thus be favored.} In this regard, pooling and sharing may constitute a fundamental mechanism for the evolution of cooperation in well-mixed fluctuating environments since it has been found that it reduces the uncertainties in future growth and, hence, brings closer the observed growth rate to the ensemble value~\cite{yaari2010cooperation,liebmann2017sharing,peters2015evolutionary}. For GBM dynamics, this has been nicely evidenced  in~\cite{peters2015evolutionary}.

Concretely, the pooling and sharing mechanism can be described as follows. A mutation introduces cooperative dynamics in a population of $N$ individuals whose resource growth is given by a GBM trajectory. In a discretized version of~(\ref{eq:gbm}), after a period of growth, the individuals pool their resources and subsequently share them \textit{equally}, resulting in the following dynamics for the resources
\begin{align}
\mathrm{d} y &= y \left( \mu \mathrm{d}t + \frac{\sigma}{\sqrt{N}} \mathrm{d}W \right).
\label{eq:gbm-pooled}
\end{align}
In~(\ref{eq:gbm-pooled}) the subscript $i$ has been dropped due to the equal sharing and $\mathrm{d}W = \frac{1}{\sqrt{N}}\sum_i \mathrm{d}W_i$ represents the pooled Wiener increment. Evidently, equation~(\ref{eq:gbm-pooled}) is a GBM with an amplitude of $\sigma/\sqrt{N}$, thus yielding a time-averaged growth of
\begin{align}
g_i(y_i(t),t) = \mu - \frac{\sigma^2}{2}\frac{1}{N}.
\end{align}

Notice that as the number of cooperating individuals increases, the time-averaged growth rate converges to the ensemble average growth. This implies that in finite populations, the introduction of new individuals always produces a net performance gain. As a result, one may conjecture that the evolution of group formation and simple multicellularity, where a class of non-cooperating unicellular species mutates to a new trait capable of forming multicellular organisms, could be a consequence of the fact that larger number of cooperators in a fluctuating environment effectively enhances the growth rate (or reduces the drift)~\cite{short2006flows,roper2013cooperatively}. Similar analogy may hold at higher levels of intelligence. As an illustration, consider situations where individuals join a community-supported agriculture to exchange their produced goods for a fixed basket of products, thereby reducing the risks in farming~\cite{adam2006community}. Another example are nations joining unions to assure sustainable economic growth through common goals~\cite{sapir2004agenda}. However, being a model of unconstrained multiplicative growth, GBM has limitations when modeling additive environments or circumstances where growth opportunities are limited due to resource or spatial constraints.

\subsection{Networked GBM}

\begin{figure}[t!]
\includegraphics[width=8.6cm]{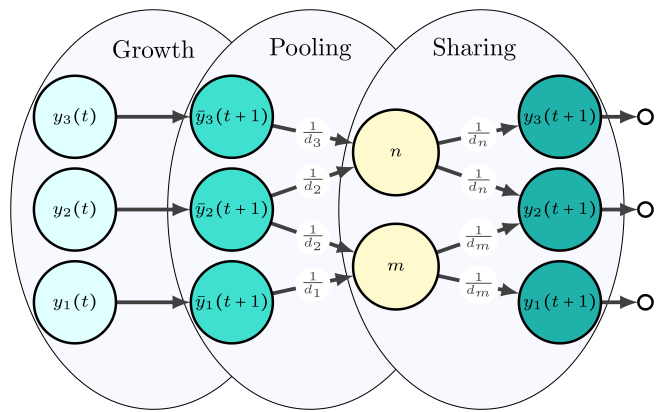}
\caption{\textbf{Networked GBM with pooling and sharing of resources.} The resources of three individuals grow according to GBM and after that they are pooled in $n$ and $m$. Finally, the pools distribute the pooled resources equally among its participants. For visualization purposes we set $\mathrm{d}t = 1$. \label{fig:model}}
\end{figure}

Real-life interactions between individuals are, however, seldom realized in a well-mixed structure, and are instead driven by a complex network of contacts~\cite{allen2017evolutionary}. To model this situation, we characterize each individual $i$ with participation in $d_i$ pools. In a discretized version of the model, each round $t$ begins with a \textit{growth phase} where the resources $y_i(t)$ of $i$ grow to $\bar{y}_i(t+\mathrm{d}t)$. The growth phase is followed by a \textit{cooperation phase} where each individual pools an equal fraction of its resources in each of the pools it belongs to. Afterwards, each pool returns an equal fraction of the pooled resources to each individual. The resulting mechanism is illustrated in Fig.~\ref{fig:model}.

The interaction structure is modeled by a connected bipartite random graph $\mathbf{B}$ between finite sets $\mathcal{N}$ of $N$ individuals and $\mathcal{M}$ of $M$ pools, with binary edge variables $B_{im} \in \left\{0,1 \right\}$ between pairs of individuals $i\in \mathcal{N}$ and pools $m\in \mathcal{M}$ ($\mathrm{B}_{im} = 1$, indicating participation of $i$ in pool $m$). The bipartite representation offers a principled way of capturing wider information regarding the group composition and network interactions~\cite{perc2013evolutionary}. In this regard, the model can be related to games of public goods played on networks, with the main difference that in our model the growth of resources of each individual \textit{precedes} the pooling phase~\cite{perc2013evolutionary,santos2008social,stojkoski2018cooperation}\footnote{It can be argued that this is more realistic for the examples of cell mutation and agricultural societies given above. In particular, cells first gather nutrients (grow), then share them. Similarly, members of community-supported agricultural first produce their goods then share them in a common pool.}.

By setting $\mathrm{d}t \to 0$, the dynamics can be explained as
\begin{align}
\mathrm{d} y_i &= \left[\sum_j^N \mathrm{A}_{ij} y_j - y_i\right] \mathrm{d}t + \sum_j^N \mathrm{A}_{ij} y_j \left( \mu \mathrm{d}t+ \sigma \mathrm{d}W_j  \right),
\label{eq:network-gbm}
\end{align}
where $\mathbf{A}$ represents a transition matrix of the network with entries $\mathrm{A}_{ij} = \sum_m^M  \frac{B_{im}}{d_m} \frac{B_{jm}}{d_j}$ determining the total allocated resources from individual $j$ to individual $i$. Equation~(\ref{eq:network-gbm}) resembles the Bouchaud--Mezard wealth reallocation model~\cite{bouchaud2000wealth,garlaschelli2008effects,berman2017empirical,ichinomiya2012wealth}, with the note that now the reallocation happens \textit{after} the growth phase.

\section{Analytical Results}
\label{sec:analytical_results}

\subsection{Time-averaged growth rate} 
For tractability, we proceed by examining a discrete version of equation~(\ref{eq:network-gbm}),
\begin{align}
y_i(t+\Delta t) = \sum_j A_{ij} y_j(t) \left[ 1 + \mu \Delta t + \sigma \varepsilon_j(t) \sqrt{\Delta t} \right],
\label{eq:network-gbm-discrete}
\end{align}
where $\varepsilon_j(t)$ is a random variable following the standard Gaussian distribution, and utilize a mean-field approach. For this purpose, we define two variables. First, the grown resources of each individual $i$ are given as 
\begin{align*}
\bar{y}_i(t+\Delta t) &= y_i(t) \left[ 1 + \mu \Delta t + \sigma \varepsilon_i(t) \sqrt{\Delta t} \right],  
\end{align*}
For large $t$ the time-averaged growth rate of this variable should be the same as $g_i(y_i(t),t)$ as its value will be dominated by $y_i(t)$.
Second, we define the mean-field around individual $i$ as the average grown resources of each of its neighbors weighted by their contributions to $i$, i.e.,
\begin{align*}
\langle \bar{y}_i \rangle = \frac{\sum_j A_{ij}\bar{y}_j}{ \sum_j A_{ij}}.  
\end{align*}
By combining the last two equations and adapting the time scale such that $\Delta t = 1$, the growth of $i$ can be approximated as
\begin{align}
g_i(y_i(t),t) &= \frac{\log(\sum_j A_{ij})}{t} + \frac{\log(\langle \bar{y}_i \rangle)}{t}.
\label{eq:discrete-growth}
\end{align}

Two implications arise from equation~(\ref{eq:discrete-growth}). First, in the transient regime there is an additive term in the growth rate which is solely dependent on the network structure. Hence, during this regime, individuals which are better connected in terms of $\sum_j A_{ij}$ should have faster growth rates. The second observation is that the second term on the right-hand side (RHS) of equation~(\ref{eq:discrete-growth}) eventually converges to the same value for each individual. This is because we study a \textit{connected} graph where participation in a pool implies that there is a path between any pair of individuals. Due to this interconnectedness, we expect that the steady state time-averaged growth of each $\langle \bar{y}_i \rangle$ will be dominated by the growth of the wealthiest individual in the network. 

The convergence of the growth rates between individuals provides a direct equivalence with the time-averaged growth rate $g(\langle y \rangle_{\mathcal{N}},t) = \frac{ \mathrm{d} \log(\langle y \rangle_{\mathcal{N}})}{\mathrm{d}t}$, which is derived from the partial ensemble average $\langle y \rangle_{\mathcal{N}}$. This object is constructed from all individuals present in the network. As a consequence, one can use It\^{o}'s lemma to directly calculate the time-averaged growth rate in the network. Formally, the lemma states that the differential of an arbitrary one-dimensional function $f(\mathbf{y},t)$ governed by an It\^{o} drift-diffusion process (such as equation~(\ref{eq:network-gbm})), is given by
\begin{align}
\mathrm{d}f(\mathbf{y},t) &= \frac{\partial f}{\partial t} \mathrm{d}t + \sum_i \frac{\partial f}{\partial y_i}\mathrm{d}y_i + \frac{1}{2}\sum_i \sum_j \frac{\partial^2 f}{\partial y_i \partial y_j} \mathrm{d}y_i \mathrm{d}y_j.
\label{eq:ito-lemma}
\end{align}
In the case of $g(\langle y \rangle_{\mathcal{N}},t)$, we have that $f(t,\mathbf{y}) = \log(\langle y \rangle_{\mathcal{N}})$. Then, the first and second derivative of $f$ with respect to $y_i$ and $y_j$ are easily calculated as $\frac{\partial f}{\partial y_i}  = \frac{1}{N} \frac{1}{\langle y \rangle_{\mathcal{N}}}$ and $\frac{\partial^2 f}{\partial y_i \partial y_j} = - \frac{1}{N^2} \frac{1}{\langle y \rangle_{\mathcal{N}}^2},$~\cite{peters2018sum}.
Moreover, this transformation makes the differential $\mathrm{d}f(\mathbf{y},t)$ ergodic, and since we are looking at steady state averages, $\mathrm{d}y_i$ and $\mathrm{d}y_i \mathrm{d}y_j$ can be substituted with their expected values $\langle \mathrm{d}y_i \rangle$ and $\langle \mathrm{d}y_i \mathrm{d}y_j \rangle$. To estimate these expectations we utilize the independent Wiener increment property $\langle \mathrm{d}W_i^2 \rangle = \mathrm{d}t$, and make use of the fact that $\sum_k A_{kj} = 1$. Further, we omit terms of order $\mathrm{d}t^2$ as they are negligible. As a result, we obtain that $\langle \mathrm{d}y_i \rangle = \left[(1+\mu) \sum_j A_{ij} y_j - y_i\right] \mathrm{d}t$
and 
$\langle \mathrm{d}y_i \mathrm{d}y_j \rangle = \sigma^2 \mathrm{d}t \sum_k A_{ik} A_{jk} y_k^2$. By inserting the estimates in equation~(\ref{eq:ito-lemma}) we can approximate the time-averaged growth rate as
\begin{align}
g(\langle y \rangle_{\mathcal{N}},t) &= \mu - \frac{\sigma^2}{2} \frac{\langle \hat{y}^2 \rangle_{\mathcal{N}}}{N},
\label{eq:network-growth-rate}
\end{align}
where $\hat{y}_i = y_i / \langle y \rangle_{\mathcal{N}}$ are the rescaled resources of individual $i$. This is a dimensionless quantity which compares the endowment of resources of an individual with the population average and as such has been particularly useful in analyses related to wealth inequality~\cite{bouchaud2000wealth}. In fact, equation~(\ref{eq:network-growth-rate}) indicates that the variance of the rescaled resources $\langle \hat{y}^2 \rangle_{\mathcal{N}}$ dictates the time-averaged growth rate. Under this model, networks with larger resource inequality, i.e. higher $\langle \hat{y}^2 \rangle_{\mathcal{N}}$, are expected to have lower steady state growth rates than those where the resources are distributed more equally. 

Additional technical details which suggest the usage of the growth rate of the partial ensemble average $g(\langle y \rangle_{\mathcal{N}}, t)$ as the growth rate of each individual is provided in the Appendix.

\subsection{Steady-state behavior}

When deriving the individual growth rate we utilized a steady state property of the system. Such properties
are key to understanding the role of complex networks within the pooling and sharing mechanism. In particular, notice that in the limit we can substitute the product of $y_j(t)$ and the exponential of~(\ref{eq:network-growth-rate}) for each $\bar{y}_j(t+ \Delta t)$, divide both sides of the equation by the population average resources and conclude that the steady state rescaled resources of individual $i$ are
\begin{align}
\lim_{t \to \infty} \hat{y}_i (t) &= v_i.
\label{eq:vi}
\end{align}
where $v_i$ is the $i$-th element of the right-eigenvector of $\mathbf{A}$ associated with the largest eigenvalue normalized in a way such that $\sum_i v_i = N$.
A direct corollary is the equilibrium individual growth rate 
\begin{align}
\lim_{t \to \infty} g_i(y_i(t),t) = \mu - \frac{\sigma^2}{2} \frac{\langle v^2 \rangle}{N}.
\label{eq:individual-growth-rate}
\end{align}
We emphasize that the quantity on the RHS of equation~(\ref{eq:individual-growth-rate}) is always greater than $\mu - \sigma^2/2$. This can be concluded by examining the optimization problem of maximizing $\langle v^2 \rangle$ constrained on $\sum_i v_i = N$,  and noting that the global maximum is always less than $N$. Therefore a network of pooling and sharing individuals on the long run will always outperform non-cooperating GBM trajectories. 
While this indicates that cooperation is a dominant trait in the population, it also asserts that, depending on the distribution of $v$, pooling and sharing may produce societies where the distribution of resources differs to a great extent from the one observed in individual trajectories~\footnote{The distribution of resources in non-cooperating GBM trajectories is log-normal}. 

\section{Numerical results}
\label{sec:numerical_results}

\subsection{Settings}
In the numerical analysis we compare the simulated dynamics of the discrete version of the networked GBM, as described with Eq.~(\ref{eq:network-gbm-discrete}), with the analytical results presented in the previous section. Due to the fact that we can only simulate for finite amount of time and as a consequence may fail to completely remove the stochasticity, we construct partial ensemble averages by averaging the results across 100 realizations of pooling and sharing.

\begin{figure*}[t!]
\includegraphics[width=16cm]{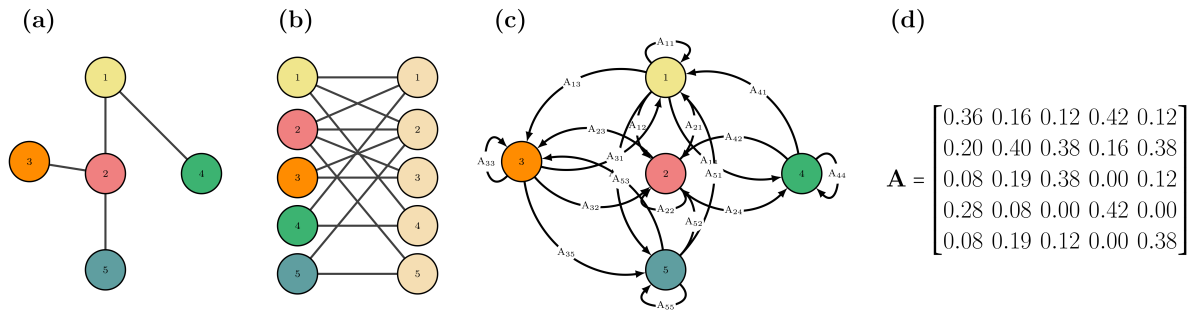}
\caption{\textbf{Graph representation.} \textbf{(a)} Example of a random graph with 5 individuals. \textbf{(b)} The  bipartite representation modeling interactions in a pooling-sharing game, as used in the numerical experiments. \textbf{(c)} The replacement graph capturing effective reallocation of resources between the individuals. The edges are the non-zero elements of the transition matrix $\mathbf{A}$, as in equation (\ref{eq:network-gbm-discrete}). \textbf{(d)} The transition matrix $\mathbf{A}$. \label{fig:graph}}
\end{figure*}

To make the analysis simpler, we formulate the interactions by considering each individual to also be a pool through which its (direct) neighbors share resources. This results in a bipartite graph where the average degree $\langle d \rangle_{\mathcal{N}}$ between individuals is equal to the average degree between pools $\langle d \rangle_{\mathcal{M}}$, i.e. $\langle d \rangle_{\mathcal{N}} = \langle d \rangle_{\mathcal{M}} = \langle d \rangle$. Fig.~\ref{fig:graph} depicts the process of mapping the original undirected random graph to a directed \textit{replacement} graph, via a bipartite graph representation which models the pooling and sharing mechanism. The edges in the replacement graph capture the elements $A_{ij}$ in the transition matrix $\mathbf{A}$.

The evaluation of the model properties is done on four types of random graphs: random d-regular graph (RR)~\cite{bollobas2013modern}, Erdos-Renyi Poisson graph (ER)~\cite{erdos1960evolution}, Watts-Strogatz small-world network (WS)~\cite{watts1998collective} and Barabasi-Albert scale-free network (BA)~\cite{barabasi1999emergence}. To capture the representative graph of each random graph that we study, for each random graph type we average the results across 100 realizations. Moreover, to distinguish the performance of the model in graphs of different size we analyze the model on both small graphs ($N = 10$) and large graphs ($N = 100$).

\subsection{Experiments}

To evaluate the performance of the model under different graph settings we conduct three experiments.

\begin{figure*}[t!]
\includegraphics[width=14cm]{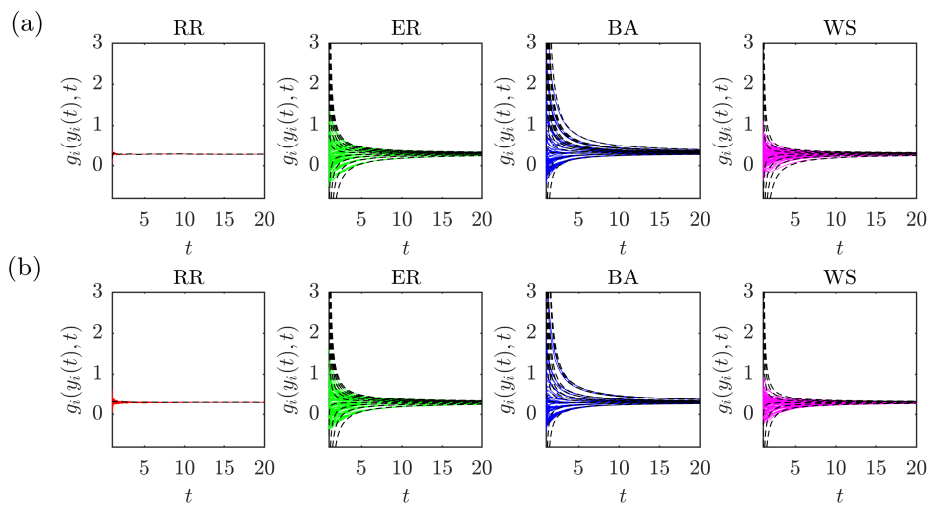}
\caption{\textbf{Transient regime dynamics.} Individual growth rate dynamics for sample RR, ER, BA and WS graphs, for \textbf{(a)} small  and \textbf{(b)} large networks. Filled lines represent simulated values while the dashed lines are the analytical solutions of the individual growth rate. The GBM parameters are set to $\mu = 0.3$ and $\sigma^2 = 0.4$. The results are averaged across 100 realizations of pooling and sharing processes with each graph each having an average degree $\langle d \rangle = 5$. \label{fig:transient_regime}}
\end{figure*}

\textit{Experiment 1.} In the first experiment we examine the transient behavior and the convergence properties of the derived growth rate as described with Eq.~(\ref{eq:discrete-growth}). The results for small and large networks are respectively given in Fig.~\ref{fig:transient_regime}a and Fig.~\ref{fig:transient_regime}b. Even though we observe that there are some discrepancies at the beginning of the simulation for each graph type an size, eventually the analytical and the numerical growth rate converge to the same value. This result holds for both small and large networks and for each random graph type, thus suggesting the plausibility of our conjecture for the convergence of the growth rate.

\begin{figure*}[t!]
\includegraphics[width=14cm]{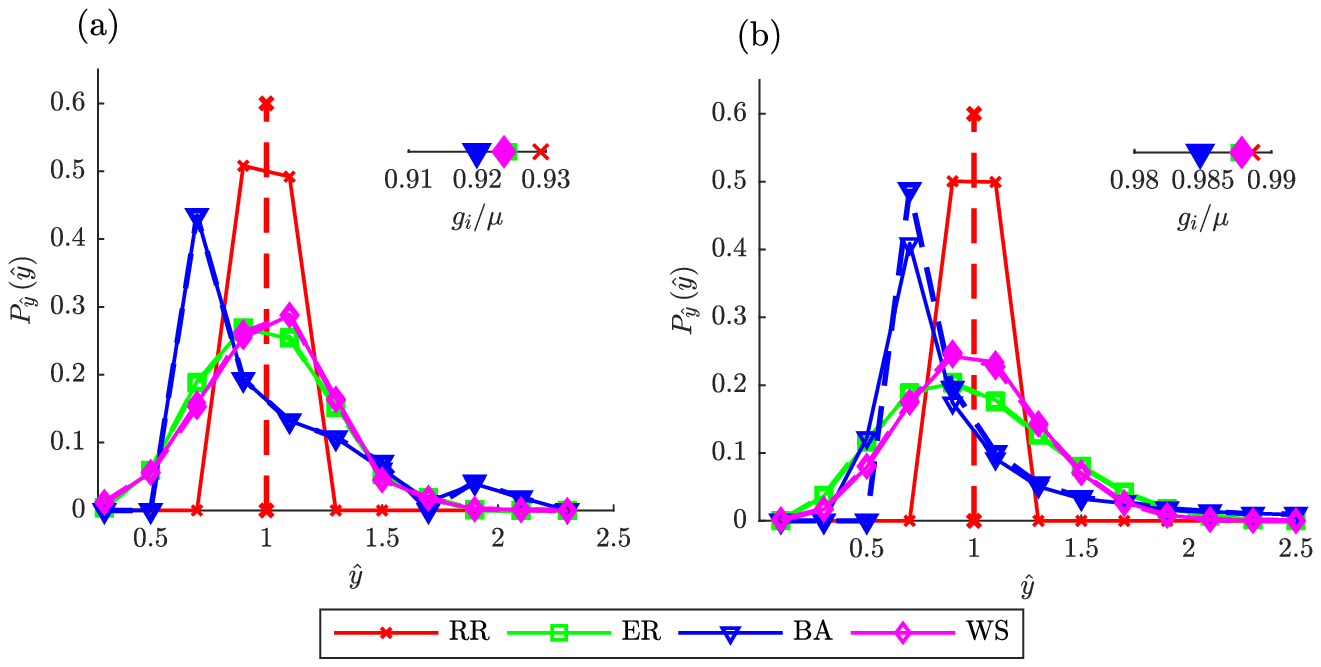}
\caption{\textbf{Steady state distribution of resources.} Estimated PDF for the rescaled resources for four different types of random graphs -- RR, ER, WS and BA, each having an average degree $\langle d \rangle = 5$. \textbf{(a)} Results for small graphs ($N = 10$). \textbf{(b)} Results for large graphs ($N = 100$). The inset plots depict the ratio of the estimated growth rate and the drift parameter for the same graphs. Filled lines represent the simulated values while the dashed lines are the analytical solutions of the corresponding variables. In the simulation $\mu = 0.3$ and $\sigma^2 = 0.4$. For each graph type, the results are averaged across $100$ realizations.  \label{fig:distribution}}
\end{figure*}

\textit{Experiment 2.} The second experiment compares the distribution of the rescaled resources in steady state, $P_\mathrm{\hat{y}} (\hat{y})$, among the graphs. Samples of the corresponding probability density functions (PDFs) are depicted in Fig.~\ref{fig:distribution}. Fig.~\ref{fig:distribution}a shows the results for small graphs while in Fig.~\ref{fig:distribution}b the corresponding results for large random graphs are provided. For both graph sizes, we notice the agreement between the analytical solution in~(\ref{eq:vi}) (the value of $v_i$) and the simulated rescaled resources, $\hat{y}_i$. In addition, independently of the graph size, we observe that the RR graph exhibits no inequality across the resources (point mass PDF), whereas the distributions of the rescaled resources in ER and WS graphs have exponential tails. Finally, the distribution of the rescaled resources in the BA graph resembles a fat tail, i.e. the resources of the individuals exhibit larger variances. As a consequence, the BA graph has the lowest steady state growth rate, followed by ER and WS, as depicted in the inset plots in Fig.~\ref{fig:distribution}. This acts as a confirmation for our second analytical finding that steady state growth rate of Eqs.~(\ref{eq:network-gbm}) and~(\ref{eq:network-gbm-discrete}) is uniquely determined by the variance of the right eigenvector associated with the largest eigenvalue of the network transition matrix $\mathbf{A}$.

\begin{figure*}[t!]
\includegraphics[width=14cm]{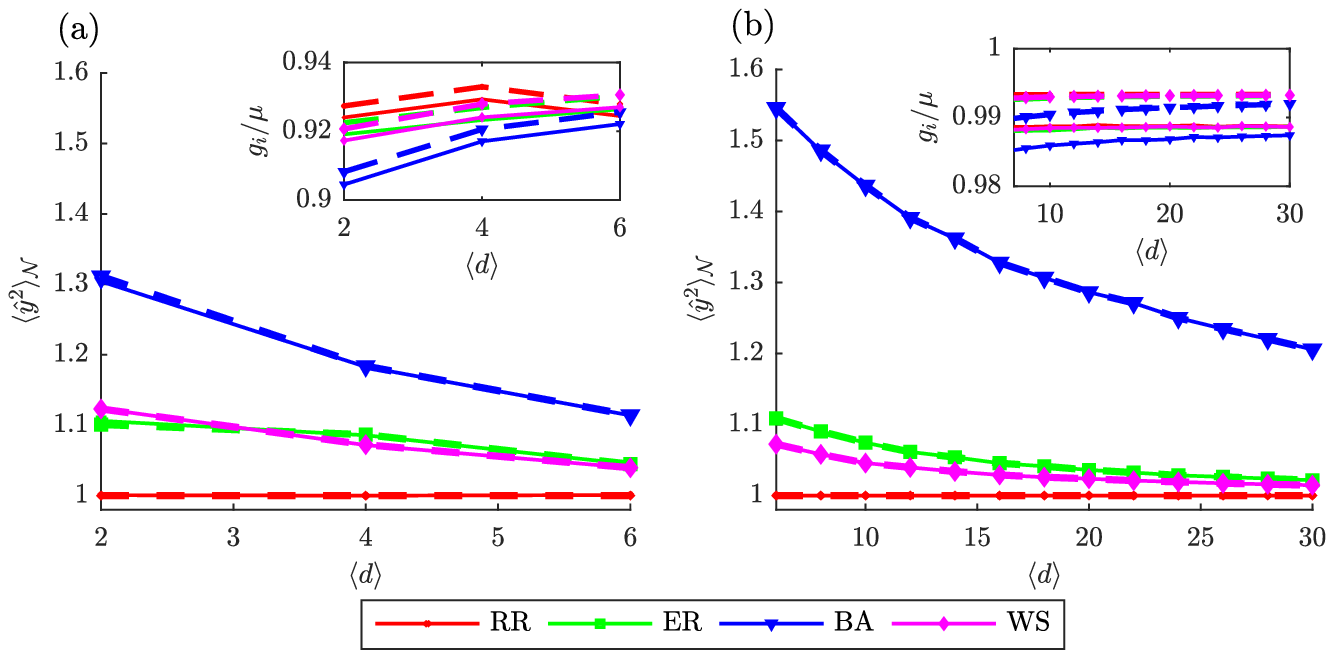}
\caption{\textbf{Network sparsity and steady state resource distribution.} The variance of rescaled resources as a function of the average degree $\langle d \rangle$ for four different types of random graphs -- RR, ER, WS and BA. \textbf{(a)} Results for small graphs ($N = 10$). \textbf{(b)} Results for large graphs ($N = 100$). The corresponding $g_i / \mu$ ratios are depicted in the insets. Filled lines represent the simulated values while the dashed lines are the analytical solutions of the corresponding variables. In the simulation $\mu = 0.3$ and $\sigma^2 = 0.4$. For each graph type, the results are averaged across $100$ realizations.  \label{fig:sparsity}}
\end{figure*}

\textit{Experiment 3.} The last experiment investigates the role of network sparsity (measured through the average degree $\langle d \rangle$), on the resource distribution. 
\textcolor{blue}{In this respect, it relates the analytical predictions described by Eq.~(\ref{eq:network-growth-rate}) with the numerical solutions of Eqs.~(\ref{eq:network-gbm}) and~(\ref{eq:network-gbm-discrete}).} Fig.~\ref{fig:sparsity} depicts the variance of rescaled resources $\langle \hat{y}^2 \rangle$ as a function of $\langle d \rangle$, for small (Fig.~\ref{fig:sparsity}a) and large graphs (Fig.~\ref{fig:sparsity}b). Moreover, the inset plots give the ratio of the individual growth rate and the drift parameter, as a function of the same variable. For both graph sizes we observe that denser ER, WS and BA graphs yield more equal resource distribution compared to their respectively sparser counterparts, whereas in the RR graph the resource distribution is invariant to the average degree. As illustrated, there is an alignment between the numerical and the theoretical results for the variance of the rescaled resources both across and within graph types. We note the slight difference between the observed (numerically obtained)  growth rate and the analytical solution which, we argue, is due to the fact that the simulations run for a finite amount of steps.

\section{Discussion}
\label{sec:discussion}
Our findings suggest that interactions on complex networks in a fluctuating environment play a critical role in the observed time-averaged growth rates and resource distribution, both in transient regime and in steady state. The cooperation dynamics is dictated by the properties of the underlying bipartite graph which models the network interactions in the pooling and sharing mechanism.

A startling example is the dynamics taking place on a BA scale-free graph, where largest discrepancies between the individual growth-rates are observed in the transient regime, as compared to ER, WS and RR graphs. Furthermore, the BA graph has the smallest time-averaged growth once the equilibrium is reached, and the most unequal resource distribution. From an evolutionary perspective, a network structure which presents with lower time-averaged growth may be interpreted as being less supportive to cooperation. It is intriguing whether there is any relationship between the apparent lower propensity to cooperation of BA scale-free networks (under the here considered interaction model) and the recent empirical evidence regarding the low-prevalence (i.e.~rarity) of scale-free networks in nature~\cite{broido2018scale,clauset2009power}.

As a takeaway, we conclude that inequality may arise as a result of the interwoven relationship between complex networks and cooperative dynamics in fluctuating environments. While it is known that certain network topologies promote inequality~\cite{barabasi1999emergence,salganik2006experimental}, the effect of cooperative behavior in structured populations is still to be determined~\cite{nishi2015inequality,chiang2015good,tsvetkova2018emergence}. As such, our investigations aim at providing deeper understanding on the nature of the relationship between these two occurrences.

Besides providing a basic model of self-reproducing living entities with temporal fluctuations, multiplicative processes are also excessively used to model self-financing
investments~\cite{peters2011optimal}, gambles~\cite{peters2016evaluating} and wealth allocation~\cite{berman2017empirical,bouchaud2000wealth}. In this respect, our findings may provide insights to economic utility theory with applications to finance, portfolio management, risk-evaluation and decision-making. In addition, they contribute to the ongoing discussions in economics and econophysics regarding the potential negative effects of wealth inequality on economic growth and development~\cite{bouchaud2000wealth,herzer2012inequality} and on the individual well-being in general.

A straightforward direction for future work is a scenario where individuals exhibit heterogeneous drifts and volatilities. There, cooperation is evolutionary advantageous only in certain parameter regimes, and thus one should investigate the dynamics under a more general model where individuals are allowed to contribute only a fraction of their resources to the pool. In this context, relations to simplistic behavioral rules that model partial cooperation, e.g.~\cite{utkovski2017promoting,stojkoski2018multiplex}, may be of particular relevance.

\section*{Acknowledgement}
This research was supported in part by DFG through grant ``Random search processes, L\'evy flights, and random walks on complex networks''.

\section{Appendix}
\label{sec:appendix}

Here we provide further mathematical logic behind our intuition to use the growth rate of the partial ensemble average $g(\langle y \rangle_{\mathcal{N}}, t)$ as the growth rate of each individual. In particular we derive two propositions which describe the dynamics of the system. The first proposition tells us that if the rescaled resources of each individual converge to a certain value, then the growth rate of every individual will also converge to the growth rate of the partial ensemble average. The second one, on the other hand, shows that if the growth rate of each individual converges to the same value, then the rescaled resources converge to $v_i$. 

\textbf{Proposition 1:} If the rescaled wealth of every individual converges to a certain real value $z_i$,  i.e. if $\lim_{t \to \infty} \hat{y}_i(t) = z_i$, with $0 < z_i < N$, then the steady state growth rate of each individual converges to the growth rate of the partial ensemble average $g(\langle y \rangle_{\mathcal{N}}, t)$.

\textit{Proof:} Suppose that $\lim_{t \to \infty} \hat{y}_i(t) = z_i$ and that the initial resources $y_i(0) = 1$ for all $i$, then
\begin{align*}
\lim_{t \to \infty} g_i( y_i(t), t) &= \lim_{t \to \infty} \frac{1}{t} \log\left(\frac{y_i(t)}{y_i(0)} \right) \\
&=  \lim_{t \to \infty} \frac{1}{t} \log \left( \langle y \rangle \cdot z_i \right) \\
&= \lim_{t \to \infty} \frac{1}{t} \log \left( \langle y \rangle \right) + \lim_{t \to \infty} \frac{1}{t} \log \left( z_i \right) \\
&= \lim_{t \to \infty} \frac{1}{t} \log \left( \langle y \rangle_{\mathcal{N}} \right) \\
&\doteq \lim_{t \to \infty} g(\langle y \rangle_{\mathcal{N}}, t).
\end{align*}

\textbf{Proposition 2:} If the steady state growth rate of each individual converges to the same value, i.e. if $\lim_{t \to \infty} g_i( y_i(t), t) = g$ for all $i$, then the steady state rescaled resources of individual $i$, $\hat{y}_i$ is given by $v_i$, where $v_i$ is the $i$-th element of the right-eigenvector of $\mathbf{A}$ associated with the largest eigenvalue and normalized in a way such that $\sum_i v_i = N$.

\textit{Proof:} Suppose that $\lim_{t \to \infty} g_i( y_i(t), t) = g$. Then by dividing equation~(\ref{eq:network-gbm-discrete}) with the average resources $\langle y(t + \Delta t)$ in period $t + \Delta t$, for the discrete version of the model it follows that:
\begin{align*}
\lim_{t \to \infty} \hat{y}_i(t + \Delta t) &= lim_{t \to \infty} \frac{\sum_j A_{ij} y_j(t) \left[ 1 + \mu \Delta t + \sigma \sqrt{\Delta t}\right]}{\left( 1 + g \right) \cdot \langle y(t) \rangle} \\
&\approx \lim_{t \to \infty} \sum_j A_{ij} \hat{y}_j (t).
\end{align*}
The last expression gives a Markov chain for which it is widely known that the stationary distribution is given by the right-eigenvector of $\mathbf{A}$ associated with the largest eigenvalue~\cite{franzke2011noise}, where its entries $v_i$ are normalized in a way such that $\sum_i v_i = N$.


%

\end{document}